\documentclass[10pt]{article}
\usepackage{amsmath}
\usepackage{graphicx}
\usepackage{subfig}
\usepackage{verbatim}
\usepackage{tikz}
\newcommand{\sign}{\text{sign}}
\usetikzlibrary{graphs}
\begin{document}

\title{A four-field gyrofluid model with neoclassical effects for the study of the rotation velocity of magnetic islands in tokamaks}
\author{A. Casolari}

\maketitle

\section{Introduction}
At equilibrium, in a tokamak, magnetic field lines lie on surfaces forming a family of nested tori, named magnetic surfaces. This structure of nested magnetic surfaces can be affected by instabilities. One of the most important ones is the so-called tearing mode, which is an instability "tearing" and reconnecting magnetic field lines. Magnetic reconnection locally breaks the topology of magnetic surfaces leading to a more energetically-favorable configuration. Magnetic islands result from the nonlinear evolution of tearing modes and represent a serious obstacle for obtaining nuclear fusion in magnetic confinement devices. In fact, the breaking of magnetic surfaces causes an increase in the heat and particle fluxes. The uncontrolled growth of magnetic islands can also lead to major disruptions, causing serious damage to the device.\\
Many efforts have been made in the past decades to develop a theory of magnetic island dynamics in tokamaks. The interest in this kind of studies is to understand the conditions for the onset of the islands in the tokamak experiments and to control them to prevent their growth to large amplitudes and the consequent negative effects on confinement. Magnetic islands arise from the nonlinear evolution of tearing modes \cite{furth1963finite}. In the presence of an equilibrium density and temperature gradient, the tearing mode acquires a propagation frequency and the instability is said a drift-tearing mode \cite{ara1978magnetic}. According to the linear drift-tearing dispersion relation, the propagation frequency of the instability should be close to the electron diamagnetic frequency, $\omega-\omega_E\approx\omega_{*e}$ \cite{coppi1965current,ara1978magnetic},  where the frequency is related to the velocity through the wave vector $\boldsymbol k$, $\omega=\boldsymbol k \cdot\boldsymbol v$. The tearing mode is an instability characterized by a long wavelength, which corresponds to a small wavevector. $\omega_E$ is the $\boldsymbol E \wedge \boldsymbol B$-drift frequency, due to the equilibrium electric field. In fact, the plasma as a whole rotates with the $\boldsymbol E \wedge \boldsymbol B$ velocity, so that this contribution must be subtracted from the island rotation velocity (Doppler shift).  Experimental observations of magnetic islands in tokamaks, under specific conditions, show a rotation frequency closer to the ion diamagnetic frequency, $\omega-\omega_E\approx\omega_{*i}$ \cite{taylor2002effect,buratti2014magnetic,buratti2016diagnostic}. This disagreement between the predictions of the linear theory and the experimental observations raises doubts on the validity of the most credited theoretical models describing magnetic island dynamics. According to recently developed models, in the presence of significant electron temperature gradients, the introduction of the so-called "mode inductivity" \cite{coppi2015open} in the Ohm's law permits the existence of modes propagating with the ion diamagnetic frequency. This effect arises naturally in the linear regime, but the experimental observations of the island rotation concern nonlinear islands, thus a direct check of the validity of this model is not currently possible. Another widely accepted interpretation of the observed rotation velocity is that, when the island width becomes larger than the ion-acoustic radius, the ion fluid cannot cross the island separatrix, thus the island is forced to propagate with the velocity of the ion flow \cite{fitzpatrick2005two}. This explanation works for islands which are large enough, but it cannot account for the transition from one diamagnetic velocity to the other. Nonlinear island dynamics is still not fully understood, and the processes that determine the island rotation velocity are under investigation.\\
Attempts to study the stationary rotation of magnetic islands have been made by Fitzpatrick \& Waelbroeck in a series of papers on the subject \cite{fitzpatrick2005two,fitzpatrick2006influence,fitzpatrick2008drift,fitzpatrick2009effect} by solving an improved version of the four-field model, previously deduced by Hazeltine, Kotschenreuther and Morrison \cite{hazeltine1985four}, which is a reduction of the two-fluid plasma description. The result of their studies is that both the island width and the neoclassical effects influence the island rotation. In particular, the critical parameters which determine the island dynamics are the ratio between the island width $w$ and the ion-acoustic radius $\rho_s$ on one side, and the ratio between the collision frequency $\nu_i$ and the bounce frequency $\omega_b$ on the other side. The first parameter determines if the island is in the sonic or hypersonic regime, which is related to the relative role of ion-acoustic waves on the flattening of the density profile inside the separatrix. The second parameter determines if the plasma is in the weak or in the intermediate damping regime, which is related to the relative strength of the neoclassical effects. The simultaneous presence of both the effects in a tokamak plasma makes it particularly difficult to determine the islands rotation velocity.\\
The purpose of this work is to investigate the magnetic island dynamics in tokamaks, in particular as regards island rotation. The attempts to study the island rotation by Fitzpatrick \& Waelbroeck rested on the inclusion of the neoclassical effects in their equations by using simplified expressions for the neoclassical  terms, together with the possibility to keep the island-size effects by using an appropriate normalization for the fields. Although their work shows results consistent with the experimental observations, their results come from a system of fluid equations which did not include the Finite Larmor Radius (FLR) corrections from the start. In this work we attempt to improve their results by starting from a set of gyrofluid equations, which result from taking the moments of the gyrokinetic equation \cite{brizard1990nonlinear,dorland1993gyrofluid}, and then reducing them to a four-field model analogous to that used by Fitzpatrick \& Waelbroeck. The neoclassical effects are included in the model by using the same simplified expressions, with an important difference. To be consistent with the gyrofluid equations, we compute the lowest order FLR corrections to the poloidal flow damping by solving the gyrokinetic equation in an appropriate limit and then computing the poloidal flow damping by following the same approach adopted in the book "Collisional transport of impurities in plasmas" by Helander \& Sigmar \cite{helander2005collisional}. The equations thus obtained have been solved by adopting a series of perturbative expansions introduced by Fitzpatrick \& Waelbroeck in their works and based on the multiple-scale approach \cite{kevorkian2012multiple}. The final equations have been solved, in two different regimes of collisionality, together with the torque balance condition, imposing that the total electromagnetic force acting on the freely-rotating islands is zero. The solution of this system of equations provides the field profiles and the self consistent phase velocity of the islands. Attempts to study both analitically and numerically the FLR effects on magnetic island evolution have been done \cite{waelbroeck2001finite,siccinio2010gyrokinetic}. In these works the focus of the authors was mainly on the analysis of the island dynamics and related phenomena, such as the emission of drift waves and the flattening of the density profile, given the island phase velocity, which was just a parameter of their models. The approach we choose, which is the same used by Fitzpatrick \& Waelbroeck, is to deduce the island rotation frequency consistently with the field profiles in a stationary regime.\\
The paper is organized as follows. In Section \ref{sec2} an analytical solution of the gyrokinetic equation is deduced and the lowest-order FLR corrections to the poloidal flow damping are calculated. In Section \ref{sec3} a four-field gyrofluid model with neoclassical effects is deduced and a series of simplifications is performed on these equations to apply them to the study of the stationary rotation of a chain of magnetic islands. In Section \ref{sec4} the torque balance condition is introduced and its explicit form is deduced for the case being considered. In Section \ref{sec5} the model is applied to the study of the weak-damping regime. In Section \ref{sec6} the model is applied to the study of the intermediate-damping regime, where a new term appears which contains the lowest-order FLR corrections to the poloidal flow damping. In Section \ref{sec7} the results of the numerical integration of the system of equations in the weak and the intermediate regime are displayed. Conclusions are drawn in Section \ref{sec8}.

\section{A particular solution of the gyrokinetic equation}
\label{sec2}
The first step in our calculation is to deduce a particular solution for the gyrokinetic equation
\begin{equation}\begin{split}
&\frac{\partial f}{\partial t}+(\boldsymbol b_0v_{\parallel}+\boldsymbol v_d)\cdot \nabla f-\left\{\frac{1}{m}\left(1+\frac{\bar{B}_{1\perp}}{B_0}\right)\left(Ze\nabla_{\parallel}\bar{\phi}_1+\mu\nabla_{\parallel}(B_0+\bar{B}_{1\parallel})\right)+\right.\\&\left.+\frac{1}{mv_{\parallel}}\boldsymbol v_d\cdot\left(Ze\nabla\bar{\phi}_1+\mu\nabla(B_0+\bar{B}_{1\parallel})\right)\right\}\frac{\partial f}{\partial v_{\parallel}}=C(f)
\label{56}
\end{split}\end{equation}
under specific simplifying hypotheses. This solution will be then used to find the FLR corrections to the neoclassical effects which occur in a tokamak. The resolution will follow the method outlined in \cite{helander2005collisional,hirshman1981neoclassical}.

\subsection{FLR expansion of the gyrokinetic equation}
By starting from Eq.\ref{56}, the distribution function $f$ is expanded in an equilibrum Maxwellian part plus a small perturbation, ordered with $\delta=\rho_i/L\ll 1$: $f=F_M+f_1$. $\rho_i$ is the ion Larmor radius, while $L$ is a macroscopic length scale. The equilibrium solution is assumed a stationary flux function, that is: $\partial F_M/\partial t=0$, $\nabla_{\parallel}F_M=0$. The following orderings are used
\begin{equation}
\frac{\partial}{\partial t}=O(\delta^2 v_{th}/L),\hspace{5mm}\frac{Ze\bar{\phi}_1}{T}=O(\delta),\hspace{5mm}\frac{v_d}{v_{th}}=O(\delta)
\label{58}
\end{equation}
With these orderings, the gyrokinetic equation Eq.\ref{56} to order $\delta$ becomes
\begin{equation}\begin{split}
&v_{\parallel}\nabla_{\parallel}f_1+\boldsymbol{v}_d\cdot\nabla(f_1+F_M)-\frac{\mu}{m}\nabla_{\parallel}B_0\frac{\partial f_1}{\partial v_{\parallel}}-\frac{1}{m}\left[Ze\nabla_{\parallel}\bar{\phi}_1+\mu\left(\nabla_{\parallel}\bar{B}_{1\parallel}+\frac{\bar{B}_{1\perp}}{B}\nabla_{\parallel}B_0\right)\right]\frac{\partial F_M}{\partial v_{\parallel}}-\\&-\frac{1}{mv_{\parallel}}\boldsymbol{v}_d\cdot[Ze\nabla\bar{\phi}_1+\mu\nabla(B_0+\bar{B}_{1\parallel})]\frac{\partial F_M}{\partial v_{\parallel}}+\frac{Ze}{m}E_{\parallel}^{(A)}\frac{\partial F_M}{\partial v_{\parallel}}=0
\end{split}\end{equation}
The term proportional to the parallel induced electric field $E_{\parallel}^{(A)}$ has been introduced to include the effect of the magnetic flux variation in a tokamak. We can meake the further assumption that $\nabla B_0\ll \nabla \phi_1,\nabla B_{1\parallel}$, meaning that the equilibrium magnetic field is almost uniform. Using the following identity to express $\boldsymbol{v}_d$ in terms of $v_{\parallel}$:
\begin{equation}
\boldsymbol v_d\cdot \nabla f=Iv_{\parallel}\nabla_{\parallel}\left(\frac{v_{\parallel}}{\Omega}\right)\frac{\partial f}{\partial \psi}
\label{57}
\end{equation}
where $I=RB_{\varphi}$ is a flux function, the gyrokinetic equation becomes:
\begin{equation}\begin{split}
&v_{\parallel}\nabla_{\parallel}\left[f_1+\frac{Iv_{\parallel}}{\Omega}\frac{\partial}{\partial \psi}(f_1+F_M)\right]=-\frac{F_M}{T}v_{\parallel}\left\{[Ze\nabla_{\parallel}\bar{\phi}_1-ZeE_{\parallel}^{(A)}+\mu\nabla_{\parallel}\bar{B}_{1\parallel}]+\right.\\&\left.+\nabla_{\parallel}\left[\frac{Iv_{\parallel}}{\Omega}\left(Ze\frac{\partial\bar{\phi}_1}{\partial \psi}+\mu\frac{\partial\bar{B}_{\parallel 1}}{\partial \psi}\right)\right]\right\}
\label{59}
\end{split}\end{equation}
Gyrokinetic theory is usually used to study turbulent transport, which is typically much larger than the  collisional one. For this reason the gyrokinetic equation we started from didn't have the collisional term on the right-hand side. To deal with neoclassical effects, we need to include the effect of collisions by using an appropriate collision operator. The contribution from $E_{\parallel}^{(A)}$ can be absorbed in a Spitzer function $f_s$, as customary in the drift-kinetic case. This one is neglected in respect to $f_1$ because, for the ions, $f_s\ll f_1$. For the fields perturbations caused by the onset of a magnetic island, the leading term is $\bar{B}_{\perp 1}$, so that we can neglect $\bar{B}_{\parallel 1}$. With these simplifications:
\begin{equation}
v_{\parallel}\nabla_{\parallel}\left[f_1+\frac{Iv_{\parallel}}{\Omega}\frac{\partial}{\partial \psi}(f_1+F_M)\right]=-\frac{F_Mv_{\parallel}}{T}\left\{Ze\nabla_{\parallel}\bar{\phi}_1+\nabla_{\parallel}\left[\frac{Iv_{\parallel}}{\Omega}\left(Ze\frac{\partial\bar{\phi}_1}{\partial \psi}\right)\right]\right\}+C(f_1)
\label{61}
\end{equation}
In the low-collisional regime, we can expand $f_1$ in a power series of the collisionality $\nu^*$, so that we can write $f_1=f_1^{(0)}+f_1^{(1)}+\cdots$ \cite{helander2005collisional,hirshman1981neoclassical}. To the two lowest orders
\begin{equation}\begin{split}
&\nabla_{\parallel}\left[f_1^{(0)}+\frac{Iv_{\parallel}}{\Omega}\frac{\partial}{\partial \psi}(f_1^{(0)}+F_M)\right]+F_M\frac{Ze}{T}\nabla_{\parallel}\left[\bar{\phi}_1+\frac{Iv_{\parallel}}{\Omega}\frac{\partial \bar{\phi}_1}{\partial \psi}\right]=0\\&
v_{\parallel}\nabla_{\parallel}\left[f_1^{(1)}+\frac{Iv_{\parallel}}{\Omega}\frac{\partial f_1^{(1)}}{\partial \psi}\right]=C(f_1^{(0)})
\label{62}
\end{split}\end{equation}
Using the fact that $\nabla_{\parallel}F_M=\nabla_{\parallel}T=0$ (neglecting significant perturbations to the temperature), the lowest order equation becomes
\begin{equation}
\nabla_{\parallel}\left[f_1^{(0)}+\frac{Iv_{\parallel}}{\Omega}\frac{\partial}{\partial \psi}(f_1^{(0)}+F_M)+F_M\frac{Ze}{T}\left(\bar{\phi}_1+\frac{Iv_{\parallel}}{\Omega}\frac{\partial \bar{\phi}_1}{\partial \psi}\right)\right]=0
\end{equation}
By integrating once, we find the following equation for $f_1^{(0)}$:
\begin{equation}
\left(1+\frac{Iv_{\parallel}}{\Omega}\frac{\partial}{\partial \psi}\right)f_1^{(0)}=g-\frac{Iv_{\parallel}}{\Omega}\frac{\partial F_M}{\partial \psi}-F_M\frac{Ze}{T}\left(1+\frac{Iv_{\parallel}}{\Omega}\frac{\partial}{\partial \psi}\right)\bar{\phi}_1
\label{63}
\end{equation}
with $g$ an unknown function such that $\nabla_{\parallel}g=0$. Eq.\ref{63} can be solved formally, by writing the solution $f_1^{(0)}$ in an integral form. Every time we deal with an equation of this form
\begin{equation}
\left(1+a\frac{d}{dx}\right)f(x)=K(x)
\label{64}
\end{equation}
the particular solution takes the form \cite{zhukovsky2012inverse}:
\begin{equation}
f(x)=\frac{e^{-x/a}}{a}\int_{x_0}^x e^{y/a}K(y)dy
\end{equation}
Eq.\ref{63} is in the form Eq.\ref{64}, so the solution for $f_1^{(0)}$ becomes
\begin{equation}
f_{1}^{(0)}=c_1(v)e^{-\psi/\psi_s}+\frac{e^{-\psi/\psi_s}}{\psi_s}\int_{\psi_0}^{\psi} e^{\chi/\psi_s}\left[g(\chi)-\psi_s\frac{\partial F_M}{\partial \chi}-Ze\frac{F_M}{T}\left(1+\psi_s\frac{\partial}{\partial\chi}\right)\bar{\phi}_1\right]d\chi
\label{65}
\end{equation}
where $\psi_s=Iv_{\parallel}/\Omega$ has the dimensions of a magnetic flux. $c_1(v)e^{-\psi/\psi_s}$ is the solution of the homogeneus equation. Once Eq.\ref{65} has been solved, we can multiply both members of the second equation of Eq.\ref{62} by $B/v_{\parallel}$ and take the flux surface average, so that we are left with the equation
\begin{equation}
\left\langle \frac{B}{v_{\parallel}}C(f_1^{(0)})\right\rangle=0
\label{66}
\end{equation}

\subsection{Analytical solution}
Tthe collision operator $C$ can be chosen in the following form \cite{helander2005collisional,hirshman1981neoclassical}:
\begin{equation}
C_{ii}(f_i)=\nu_D^{ii}(v)\left(\mathcal{L}(f_{i1})+\frac{m_iv_{\parallel}u_i}{T_i}f_{Mi}\right)
\label{69}
\end{equation}
By introducing the following definition of the Lorentz operator
\begin{equation}
\mathcal{L}=\frac{2hv_{\parallel}}{v^2}\frac{\partial}{\partial\lambda}\lambda v_{\parallel}\frac{\partial}{\partial\lambda}
\label{lorentz_operator}
\end{equation}
where $h\equiv B_0/B$ is the toroidal metric coefficient and $\lambda$ is related to the particles pitch angle by $\lambda\equiv h\sin^2\alpha$, Eq.\ref{66} becomes:
\begin{equation}\begin{split}
&\left\langle B\left\{\frac{2h}{v^2}\frac{\partial}{\partial \lambda}\lambda v_{\parallel}\frac{\partial}{\partial \lambda}\left(g-e^{-\psi/\psi_s}\int_{\psi_0}^{\psi}d\chi e^{\chi/\psi_s}\partial_{\chi}g+e^{-\psi/\psi_s}c_1\right)+\frac{I}{\Omega}\left[\partial_{\psi}F_M+\right.\right.\right.\\&\left.\left.\left.+Ze\frac{F_M}{T}\partial_{\psi}\bar{\phi}_1-e^{-\psi/\psi_s}\int_{\psi_0}^{\psi}d\chi e^{\chi/\psi_s}\left(\partial^2_{\chi}F_M+Ze\partial_{\chi}\left(\frac{F_M}{T}\partial_{\chi}\bar{\phi}_1\right)\right)\right]+\frac{m_iu_i}{T}F_M\right\}\right\rangle=0
\end{split}\end{equation}
We define the auxiliary function $J$:
\begin{equation}
J=g-e^{-\psi/\psi_s}\int_{\psi_0}^{\psi}d\chi e^{\chi/\psi_s}\partial_{\chi}g+e^{-\psi/\psi_s}c_1
\end{equation}
The equation for $J$ is
\begin{equation}\begin{split}
&\frac{\partial}{\partial \lambda}\lambda \left\langle v_{\parallel}\right\rangle\frac{\partial}{\partial \lambda}J=-\frac{v^2}{2}\left\{\frac{I}{h\Omega}\left[\partial_{\psi}F_M+Ze\frac{F_M}{T}\partial_{\psi}\bar{\phi}_1-\right.\right.\\&\left.\left.-e^{-\psi/\psi_s}\int_{\psi_0}^{\psi}d\chi e^{\chi/\psi_s}\left(\partial^2_{\chi}F_M+Ze\partial_{\chi}\left(\frac{F_M}{T}\partial_{\chi}\bar{\phi}_1\right)\right)\right]+\left\langle\frac{u_i}{h}\right\rangle\frac{m_i}{T}F_M\right\}
\label{72}
\end{split}\end{equation}
From the form of Eq.\ref{72}, we can deduce that $J$ plays the role of the function which, in the drift-kinetic equation, vanishes in the trapped particle space, so that the solution of Eq.\ref{72} is
\begin{equation}\begin{split}
&J=H(\lambda_c-\lambda)\frac{v^2}{2}\int_{\lambda}^{\lambda_c}\frac{d\lambda'}{\left\langle v_{\parallel}(\lambda') \right\rangle}\left\{\frac{I}{h\Omega}\left[\partial_{\psi}F_M+Ze\frac{F_M}{T}\partial_{\psi}\bar{\phi}_1-\right.\right.\\&\left.\left.-e^{-\psi/\psi_s}\int_{\psi_0}^{\psi}d\chi e^{\chi/\psi_s}\left(\partial^2_{\chi}F_M+Ze\partial_{\chi}\left(\frac{F_M}{T}\partial_{\chi}\bar{\phi}_1\right)\right)\right]+\left\langle \frac{u_i}{h}\right\rangle\frac{m_i}{T}F_M\right\}
\end{split}\end{equation}
where $H$ is the Heaviside function. By using a few results from drift-kinetik theory, we find the following solution:
\begin{equation}\begin{split}
&f_1^{(0)}=-\frac{Iv_{\parallel}}{\Omega}\left[\partial_{\psi}F_M+Ze\frac{F_M}{T}\partial_{\psi}\bar{\phi}_1-e^{-\psi/\psi_s}\int_{\psi_0}^{\psi}d\chi e^{\chi/\psi_s}\left(\partial^2_{\chi}F_M+Ze\partial_{\chi}\left(\frac{F_M}{T}\partial_{\chi}\bar{\phi}_1\right)\right)\right]+\\&+\frac{IHV_{\parallel}}{h\Omega}\left(\frac{mv^2}{2T}-1.33\right)\frac{d\log T}{d\psi}F_M-\frac{Ze\bar{\phi}_1}{T}F_M+e^{-\psi/\psi_s}\int_{\psi_0}^{\psi}d\chi e^{\chi/\psi_s}Ze\partial_{\chi}\left(\frac{F_M}{T}\bar{\phi}_1\right)-\\&-\frac{IHV_{\parallel}}{h\Omega}e^{-\psi/\psi_s}\int_{\psi_0}^{\psi} d\chi e^{\chi/\psi_s}\left[\partial^2_{\chi}F_M+Ze\partial_{\chi}\left(\frac{F_M}{T}\partial_{\chi}\bar{\phi}_1\right)\right]-\\&-\frac{IHV_{\parallel}}{h\Omega}\frac{F_M}{\left\{\nu_D^{ii}\right\}}\left\{\nu_D^{ii}e^{-\psi/\psi_s}\int_{\psi_0}^{\psi} d\chi e^{\chi/\psi_s}\left[\frac{\partial^2_{\chi}F_M}{F_M}+\frac{Ze}{F_M}\partial_{\chi}\left(\frac{F_M}{T}\partial_{\chi}\bar{\phi}_1\right)\right]\right\}
\label{76}
\end{split}\end{equation}
The curly braces in the last line of Eq.\ref{76} represent the velocity-space average, which is defined as:
\begin{equation}
\{F\}=\int d^3v F\frac{mv^2}{nT}F_M
\end{equation}
The solution Eq.\ref{76} still contains terms in an integral form. However, it contains the FLR corrections which provide, after velocity-space integration, the modified transport coefficients in the different collisionality regimes.

\subsection{Poloidal flow damping}
Eq.\ref{76} can be used to compute the neoclassical effects, in particular the poloidal flow damping, which comes from the toroidal geometry, through the equation \cite{helander2005collisional,hirshman1981neoclassical}
\begin{equation}
\left\langle\boldsymbol B\cdot\nabla\cdot\boldsymbol \pi\right\rangle=\left\langle B(F_{\parallel}+nZeE_{\parallel}^{(A)})\right\rangle
\label{77}
\end{equation}
where $\boldsymbol\pi$ is the stress tensor and $F_{\parallel}$ is the parallel component of the friction force, which is defined in terms of the distribution function
\begin{equation}
F_{\parallel}\equiv\int mv_{\parallel}C(f_1)d^3v
\label{78}
\end{equation}
We can now use the Spitzer function to eliminate the term proportional to the inductive electric field and remind that, for the ions, the Spitzer function is negligible in respect to the function $f_1$. Using the particular form for the collision operator Eq.\ref{69}, together with Eq.\ref{78}, Eq.\ref{77} becomes
\begin{equation}
\left\langle\boldsymbol B\cdot\nabla\cdot\boldsymbol \pi\right\rangle=\left\langle B\int mv_{\parallel}\nu_D^{ii}\left(\mathcal{L}(f_1^{(0)})+\frac{mv_{\parallel}U_{\parallel i}}{T}F_M\right)d^3v\right\rangle
\label{79}
\end{equation}
$U_{\parallel i}$ is the parallel flow velocity of the ions, which is defined as
\begin{equation}
U_{\parallel i}=\int v_{\parallel}f_1^{(0)}d^3v
\label{80}
\end{equation}
Eq.\ref{76} contains the FLR effects in terms of integral expressions. Such quantities can be expanded in a power series in respect to $\psi_s$ performing an integration by parts
\begin{equation}
e^{-\psi/\psi_s}\int d\chi e^{\chi/\psi_s} F=\psi_sF-\psi_s^2\partial_{\psi}F+O(\psi_s^3\partial^2_{\psi}F)
\label{81}
\end{equation}
This expansion is made possible by the smallness of the ion Larmor radius: in fact, after velocity integration, $\psi_s/\psi=\rho_i/L\ll 1$. From this result we notice that, when taking the velocity moments of the distribution function, only the terms which have the correct parity will remain and the others will be zero. This is particularly important because, from the lowest order expansion Eq.\ref{81}, only terms proportional to $\rho_i^2$ will remain. When applying this expansion to Eq.\ref{76} and applying it to Eq.\ref{80}, we find:
\begin{equation}\begin{split}
&U_{\parallel}=-\frac{I}{\Omega}\int d^3vv_{\parallel}^2\left(\partial_{\psi}F_M+Ze\frac{F_M}{T}\partial_{\psi}\bar{\phi}_1\right)+\frac{I}{\Omega}\int d^3vv_{\parallel}^2\left(\frac{mv^2}{2T}-1.33\right)\frac{d\log T}{d\psi}F_M+\\&+\frac{I}{\Omega}\int d^3vv_{\parallel}^2Ze\partial_{\psi}\left(\frac{F_M}{T}\bar{\phi}_1\right)+\frac{I^3}{\Omega^3}\int d^3v \frac{v_{\parallel}^2F_M}{\{\nu_D^{ii}\}}\left\{\nu_D^{ii}v_{\parallel}^2\left[\frac{\partial^3_{\psi}F_M}{F_M}+\frac{Ze}{F_M}\partial^2_{\psi}\left(\frac{F_M}{T}\partial_{\psi}\bar{\phi}_1\right)\right]\right\}
\label{82}
\end{split}\end{equation}
Since the radial derivatives are steep, we only keep the highest order derivatives in Eq.\ref{82};
\begin{equation}
\partial^3_{\psi}F_M+Ze\partial^2_{\psi}\left(\frac{F_M}{T}\partial_{\psi}\bar{\phi}_1\right)=\left[\frac{1}{P}\frac{d^3 P}{d\psi^3}+2\frac{Ze}{T}\frac{d^3\bar{\phi}_1}{d\psi^3}+\left(\frac{mv^2}{2T}-\frac{5}{2}\right)\frac{1}{T}\frac{d^3T}{d\psi^3}\right]F_M
\label{84}
\end{equation}
Inserting the solution Eq.\ref{76} expanded according to Eq.\ref{81} into Eq.\ref{79}, and by using Eq.\ref{84}:
\begin{equation}\begin{split}
\left\langle\boldsymbol B\cdot\nabla\cdot\boldsymbol \pi\right\rangle\approx & B\mu_{01}m_inf_t\nu_{ii}\frac{IT_i}{m_i\Omega_i}\left\{1.17\frac{d\log T_i}{d\psi}+\right.\\&\left.+\frac{I^2T_i}{m_i\Omega_i^2}\left[2.70\left(\frac{1}{P}\frac{d^3P}{d\psi^3}+2\frac{Ze}{T}\frac{d^3\bar{\phi}_1}{d\psi^3}\right)-0.70\frac{1}{T}\frac{d^3T}{d\psi^3}\right]\right\}
\label{85}
\end{split}\end{equation}
where $\mu_{0i}=\{\nu_D^{ii}\}$, $f_t$ is the fraction of trapped particles and the following properties have been used \cite{helander2005collisional}:
\begin{equation}
\{\nu_D^{ii}\}\approx 0.53\hspace{5mm}\{\nu_D^{ii}x^2\}\approx 0.71\hspace{5mm}\{\nu_D^{ii}x^4\}\approx 1.59
\end{equation}
where $x^2=v^2/v_{th}^2$. The first term in Eq.\ref{85} is the result from drift-kinetic theory. The additional terms are the first order FLR corrections, which are proportional to $\rho_i^2$.

\section{Four-field gyrofluid model}
\label{sec3}
The system of gyrofluid equations originally developed by P. B. Snider \cite{snyder1999gyrofluid} consists of six equations, evolving the density, the parallel velocity, the parallel and perpendicular pressure and the parallel and perpendicular heat flux for each particle species. Here we just need the first two of them, together with the vorticity equation, which can be deduced from the quasi-neutrality condition
\begin{equation}
n_e=\Gamma_0^{1/2}n_i+n_0\frac{Ze}{T_i}(\Gamma_0-1)\phi
\label{quasi_neutrality}
\end{equation}
where $\Gamma_0=\left\langle J_0^2\right\rangle$ is the velocity average of the zero-order Bessel function, with argument $k_{\perp}^2\rho_i^2$, which comes from the gyroaverage involved in the gyrocenter transformation. The second term on the right-hand side of Eq.\ref{quasi_neutrality} is the so-called polarization density. The momentum equation for the two species electrons and ions can be written as:
\begin{equation}\begin{split}
&\frac{\partial \bar{n}_i}{\partial t}+\boldsymbol v_{\Phi}\cdot \nabla \bar{n}_i+n_0\bar{\nabla}_{\parallel}\bar{u}_{\parallel i}=0\\&
\frac{\partial n_e}{\partial t}+\boldsymbol v_{\phi}\cdot \nabla n_e+n_0\nabla_{\parallel}u_{\parallel e}=0\\&
m_in_0\left(\frac{\partial \bar{u}_{\parallel i}}{\partial t}+\boldsymbol v_{\Phi}\cdot \nabla \bar{u}_{\parallel i}\right)=-T_{0i}\bar{\nabla}_{\parallel}\bar{n}_i+en_0\left(\frac{\partial \Psi}{\partial t}-\nabla_{\parallel}\Phi\right)+F_{ie}\\&
m_en_0\left(\frac{\partial u_{\parallel e}}{\partial t}+\boldsymbol v_{\phi}\cdot \nabla u_{\parallel e}\right)=-T_{0e}\nabla_{\parallel}n_e-en_0\left(\frac{\partial \psi}{\partial t}-\nabla_{\parallel}\phi\right)+F_{ei}
\label{cont_momentum_simp}
\end{split}\end{equation}
where $\Phi=\Gamma_0^{1/2}\phi$, $\Psi=\Gamma_0^{1/2}\psi$ are the gyroaveraged fields, $F_{ie}$ and $F_{ei}$ are the collisional friction forces, whose parallel component is defined in Eq.\ref{78}. $\bar{\nabla}_{\parallel}$ is the parallel gradient performed along the gyroaveraged magnetic field. $\bar{n}_i$ and $\bar{u}_{\parallel i}$ are the ion density and parallel velocity expressed in the gyrocenter coordinates. Owing to momentum conservation in Coulomb collisions, the property $F_{ie}=-F_{ei}$ holds. We used $P=nT_0$, with $T_0$ uniform and constant. The common equilibrium density $n_0$ multiplies the electric force term in the momentum equations because the electric field is perturbative. By proceeding similarly to the quasi-neutrality calculation, we can find the gyrokinetic definition of the current:
\begin{equation}
J_{\parallel}=-en_0(u_{\parallel e}-\Gamma_0^{1/2}\bar{u}_{\parallel i})
\label{current_def}
\end{equation}
Introducing the Debye length $\lambda_{Di}=\sqrt{T_i/(n_ie^2)}$, Eq.\ref{quasi_neutrality} becomes:
\begin{equation}
\frac{1}{\lambda_{Di}^2}(\Gamma_0-1)\phi=e(n_e-\Gamma_0^{1/2}\bar{n}_i)
\label{quasi_neut_aagain}
\end{equation}
Taking the time derivative of Eq.\ref{quasi_neut_aagain} and using the equations above, we obtain the following vorticity equation:
\begin{equation}
\left(\frac{\partial}{\partial t}+\boldsymbol v_{\phi}\cdot\nabla\right)\left(\frac{1}{\lambda_{Di}^2}(\Gamma_0-1)\phi+e(\Gamma_0^{1/2}-1)\bar{n}_i\right)=\nabla_{\parallel}J_{\parallel}+e(\Gamma_0^{1/2}-1)\boldsymbol v_{\phi}\cdot \nabla \bar{n}_i
-e\nabla_{\parallel}(\Gamma_0^{1/2}-1)\bar{u}_{\parallel i}-e(\bar{\nabla}_{\parallel}-\nabla_{\parallel})\bar{u}_{\parallel i}
\label{gyro_vorticity}
\end{equation}
We can further simplify this system of equations by neglecting the electron inertia in the electron momentum equation, which becomes the generalized Ohm law. Then we use the quasi-neutrality condition Eq.\ref{quasi_neut_aagain} to express $n_e$ in terms of $\bar{n}_i$ and we sum the momentum equations of the ions and the electrons.\\
The system of equations we get is:
\begin{equation}\begin{split}
&\frac{\partial \bar{n}_i}{\partial t}+\boldsymbol v_{\Phi}\cdot \nabla \bar{n}_i+n_0\bar{\nabla}_{\parallel}\bar{u}_{\parallel i}=0\\&
\frac{\partial \psi}{\partial t}-\nabla_{\parallel}\phi+\frac{T_{0e}}{en_0}\nabla_{\parallel}\left[\Gamma_0^{1/2}\bar{n}_i+n_0(\Gamma_0-1)\frac{e\phi}{T_{0i}}\right]=\eta J_{\parallel}\\&
m_i\left(\frac{\partial}{\partial t}+\boldsymbol v_{\Phi}\cdot \nabla \right)\bar{u}_{\parallel i}=-\frac{T_{0e}}{n_0}\bar{\nabla}_{\parallel}\left[(\tau+\Gamma_0^{1/2})\bar{n}_i+(\Gamma_0-1)\frac{e\phi}{T_{0i}}\right]\\&
\left(\frac{\partial}{\partial t}+\boldsymbol v_{\phi}\cdot\nabla\right)\left(\frac{1}{\lambda_{Di}^2}(\Gamma_0-1)\phi+e(\Gamma_0^{1/2}-1)\bar{n}_i\right)=\nabla_{\parallel}J_{\parallel}+e(\Gamma_0^{1/2}-1)\boldsymbol v_{\phi}\cdot \nabla \bar{n}_i
\label{gyro_system}
\end{split}\end{equation}
where $\tau=T_{0i}/T_{0e}$. We neglected in the parallel momentum equation the electric force coming from the difference between the fields $\psi$, $\phi$ and their gyroaverage. $\boldsymbol v_{\Phi}$ and $\bar{\nabla}_{\parallel}$ are the $E\wedge B$ velocity and the parallel gradient on the fields calculated with the gyroaveraged fields: $B_0 \boldsymbol v_{\Phi}\cdot\nabla=(\hat{e}_z\wedge\nabla\Gamma_0^{1/2}\phi)\cdot\nabla$, $B_0 \bar{\nabla}_{\parallel}=(\hat{e}_z\wedge\nabla\Gamma_0^{1/2}\psi)\cdot\nabla$.\\
The quantity $1/\lambda_{Di}^2(\Gamma_0-1)\phi+e(\Gamma_0^{1/2}-1)\bar{n}_i$ is the gyrokinetic vorticity. In the limit of large wavelengths, this quantity reduces to $\rho_i^2e(n_oe/T_i\nabla_{\perp}^2\phi+1/2\nabla_{\perp}^2 n_i)$. The first term corresponds to the $E\wedge B$ drift velocity, while the second one represents the contribution from the diamagnetic velocity. The factor $1/2$ appearing in front of this term comes from the expansion of the gyroverage operator \cite{xu2013gyro}. Attempts to study an Hamiltonian version of these equations, both analytically and numerically, has been done by different authors \cite{comisso2012numerical,waelbroeck2012compressible}.\\
Gyrofluid equations surpass the fluid equations because they include the FLR effects which come from the gyrokinetic theory. However, these FLR effects are present in the form of nonlinear differential operators, quite difficult to deal with both analytically and numerically. Several attempts have been made by different authors to deal with these operators by approximating them with power expansions and elementary functions. An overview of these attempts is provided in  \cite{dorland1993gyrofluid}. The operators $\Gamma_0$ and $\Gamma_0^{1/2}$ involve all the even powers of $b=k_{\perp}^2\rho_i^2$:
\begin{equation}\begin{split}
&\Gamma_0=\left\langle J_0^2\right\rangle =I_0(b)e^{-b}=1-b+O(b^2)\\&
\Gamma_0^{1/2}\approx\left\langle J_0^2\right\rangle^{1/2} = I_0^{1/2}(b)e^{-b/2}=1-b/2+O(b^2)
\end{split}\end{equation}
where $I_0$ is the modified Bessel function. The Taylor expansion of these operators provides the FLR corrections to all orders in $b$. In the limit of large wavelengths $k_{\perp}^2\rho_i^2\ll 1$ the power expansion can be truncated to a low order (usually the second order is already a good approximation). However, in the limit of small wavelengths $k_{\perp}^2\rho_i^2\gg 1$ (or more realistically, $k_{\perp}^2\rho_i^2=O(1)$), the power expansion isn't a good approximation any longer.\\
If we introduce the following normalization for the fields \cite{fitzpatrick2009effect}:
\begin{equation}\begin{split}
&\hat{\psi}=\frac{L_s}{B_0w^2}\psi,\hspace{5mm}\hat{n}=-\frac{L_n}{w}\frac{\bar{n}_i}{n_0},\hspace{5mm}\hat{\phi}=-\frac{\phi}{wV_{*e}B_0}+\hat{x}V_p\\&\hat{V}=\frac{\epsilon}{q}\frac{V_{\parallel i}}{V_{*e}},\hspace{5mm}\hat{J}=\frac{L_S\mu_0}{B_0\delta_e}J_{\parallel},\hspace{5mm}\hat{\eta}=\frac{\eta}{\mu_0k_{\theta}V_{*e}w^2}
\label{normaliz_fields}
\end{split}\end{equation}
where $w$ is the island width, $\delta_e=\beta/\alpha^2$, $\alpha^2=w^2/\rho_s^2L_n^2/L_s^2$ and $\rho_s=c_s/\Omega_i$ is the ion-acoustic radius. The $x$-derivatives are normalized to $w$, the $y$-derivatives are normalized to $L_n=1/k_{\theta}$ and the time derivatives are normalized to $k_{\theta}V_{*e}$. $V_p$ is the unknown island rotation velocity. The additional term $\hat{x}V_p$ in the normalized electrostatic potential represents the contribution from the island-induced electric field. With this choice for the normalization, the gradients length-scale of the fields in the radial direction is comparable with the island width $w$.

\subsection{Neoclassical effects}
The neoclassical effects come from the inhomogeneity of the magnetic field and the low collisionality of the plasma. Neoclassical theory and the poloidal flow damping have been thoroughly described in \cite{hirshman1981neoclassical,helander2005collisional}. In addition to the poloidal damping caused by the toroidal shape of the tokamak, there is a similar phenomenon caused by non-axisymmetric effects, such as the magnetic islands. The broken poloidal symmetry of the torus causes the travelling particles to experience a magnetic-mirror effect, which leads to the phenomenon of banana orbits and the consequent poloidal flow damping. Analogously, the broken axisymmetry caused by the presence of magnetic islands leads to a situation of "helically-trapped particles", causing a braking effect on the plasma rotation called "island-induce flow damping" \cite{shaing2004plasma,shaing2011theory}.
The procedure to obtain the non-axisymmetric effects on the plasma rotation is analogous to that used in the axisymmetric case but the calculations are much more involved because of the complex shape of the flux surfaces. The island-induced flow damping is proportional to the island width $w$ squared \cite{shaing2004plasma,fitzpatrick2012spontaneous}, so that its effect becomes significative only when the FLR effects are negligible. For this reason we chose not to compute the FLR corrections to this term by solving the gyrokinetic equation. The only correction we are going to keep is the usual lowest-order expansion of the gyroaveraged electrostatic potential. That said, let us consider the divergence of the ion stress tensor we deduced in the section above: according to \cite{fitzpatrick2012spontaneous}, the flow damping can be included in the system of equations we are using by imposing that, under the effect of this damping, the poloidal rotation velocity tends to relax to its neoclassical value. This amounts to introducing the following damping term:
\begin{equation}
m_in_i\nu_{\theta}(\boldsymbol V_i\cdot \boldsymbol e_{\theta}-V_{\theta}^{nc})
\label{damping_model}
\end{equation}
After switching from $\psi$ to $r$, simplifying a few terms and introducing the notation $T'=dT/dr$, Eq.\ref{damping_model} becomes:
\begin{equation}
m_in_i\nu_{\theta}\left\{V_{\parallel i}\hat{b}\cdot\boldsymbol e_{\theta}+\frac{1}{B_0}\bar{\phi}_1'+\frac{T_i}{en_0B_0}\bar{n}_i'(1-c_{\theta})-\frac{T_i}{en_0B_0}\left[c_1\frac{n_i'''}{n_0}+c_2\frac{Ze}{T_i}\bar{\phi}_1'''\right]\right\}
\label{poloidal_damp_term}
\end{equation}
where $c_{\theta}=1.17\eta_i$, $c_1=2.70+2\eta_i$, $c_2=5.40$ and $\eta_i=L_n/L_T$ is the ratio between the length scales of the density and temperature gradients. If we use the normalization for the fields introduced in Eq.\ref{normaliz_fields}, we find the following adimensional expression:
\begin{equation}
-\hat{\nu}_{\theta}\left\{\hat{V}+V_p-\partial_{\hat{x}}[\Gamma_0^{1/2}\hat{\phi}+\tau\hat{n}(1-c_{\theta})]+\tau\rho^2\partial_{\hat{x}}[c_1\partial^2_{\hat{x}}\hat{n}+c_2\partial^2_{\hat{x}}\hat{\phi}]\right\}
\end{equation}
where $\hat{\nu}_{\theta}=\nu_{\theta}/(k_{\theta}V_{*e})$ is the poloidal damping coefficient, which is determined by the kinetic theory. An analogous expression exists for the island-induced perpendicular flow damping \cite{shaing2004plasma,fitzpatrick2012spontaneous}. When using the normalization Eq.\ref{normaliz_fields}, it takes the following form:
\begin{equation}
\hat{\nu}_{\perp}\partial_{\hat{x}}[\Gamma_0^{1/2}\hat{\phi}+\tau\hat{n}(1-c_{\perp})]
\label{perp_damp_adim1}
\end{equation}
$\hat{\nu}_{\perp}=\nu_{\perp}/(k_{\theta}V_{*e})$ is the perpendicular damping coefficient, which is proportional to $w^2$, and $c_{\perp}=2.37$. As emphasized in \cite{fitzpatrick2012spontaneous}, the island-induced flow damping acts in the perpendicular direction, so that it doesn't contribute to the parallel momentum equation. When including these effects in Eq.\ref{gyro_system}, the normalized system of equations becomes:
\begin{equation}\begin{split}
&\frac{\partial \hat{n}}{\partial \hat{t}}=[\hat{\phi},\hat{n}]+[\hat{V},\hat{\psi}]+\frac{\rho^2}{2}\{[\partial^2_{\hat{x}}\hat{\phi},\hat{n}]+[\hat{V},\partial^2_{\hat{x}}\hat{\psi}]\}\\&
\frac{\partial \hat{\psi}}{\partial \hat{t}}=[\hat{\phi}-\hat{n},\hat{\psi}]+\hat{\eta}\delta_e\hat{J}-\rho^2[\partial^2_{\hat{x}}\hat{\phi}/\tau+\partial^2_{\hat{x}}\hat{n}/2,\hat{\psi}]\\&
\frac{\partial \hat{V}}{\partial \hat{t}}=[\hat{\phi},\hat{V}]+\alpha^2(1+\tau)[\hat{n},\hat{\psi}]+\rho^2\left\{\frac{1}{2}[\partial^2_{\hat{x}}\hat{\phi},\hat{V}]+\frac{\alpha^2}{2}\left((1+\tau)[\hat{n},\partial^2_{\hat{x}}\hat{\psi}]+[\partial^2_{\hat{x}}\hat{n},\hat{\psi}]\right)+\frac{1}{\tau}[\hat{\psi},\partial^2_{\hat{x}}\hat{\phi}]\right\}-\\&-\hat{\nu}_{\theta}\left(\frac{\epsilon}{q}\right)^2\left\{\hat{V}+V_p-\partial_{\hat{x}}\left[\left(1+\frac{\rho^2}{2}\partial^2_{\hat{x}}\right)\hat{\phi}+\tau\hat{n}(1-c_{\theta})-\tau c_1\rho^2\partial^2_{\hat{x}}\hat{n}-\tau c_2\rho^2\partial^2_{\hat{x}}\hat{\phi}\right]\right\}\\&
\frac{\partial }{\partial \hat{t}}\partial^2_{\hat{x}}\left(\hat{\phi}+\frac{\tau}{2}\hat{n}\right)=\left[\hat{\phi},\partial^2_{\hat{x}}\left(\hat{\phi}+\frac{\tau}{2}\hat{n}\right)\right]+\frac{\tau}{2}[\hat{n},\partial^2_{\hat{x}}\hat{\phi}]+[\hat{J},\hat{\psi}]+\rho^2\left\{\left[\hat{\phi},\partial^4_{\hat{x}}\left(\hat{\phi}+\frac{\tau}{4}\hat{n}\right)\right]+\frac{\tau}{4}[\hat{n},\partial^4_{\hat{x}}\hat{\phi}]\right\}-\\&-\nu_{\perp}\partial^2_{\hat{x}}\left[\left(1+\frac{\rho^2}{2}\partial^2_{\hat{x}}\right)\hat{\phi}+\tau\hat{n}(1-c_{\perp})\right]+\nu_{\theta}\partial_{\hat{x}}\left\{\hat{V}-\partial_{\hat{x}}\left[\left(1+\frac{\rho^2}{2}\partial^2_{\hat{x}}\right)\hat{\phi}+\tau\hat{n}(1-c_{\theta})-\tau c_1\rho^2\partial^2_{\hat{x}}\hat{n}-\tau c_2\rho^2\partial^2_{\hat{x}}\hat{\phi}\right]\right\}
\label{normaliz_system_bis}
\end{split}\end{equation}
Note that the FLR corrections coming from the analytical resolution of the gyrokinetic equation are consistent with those coming from the the small-Larmor-radius expansion of the gyrofluid equations. In fact, although the perturbation to the distribution function $f_1$ was assumed ordered with $\delta=\rho_i/L\ll 1$, the distinction between the parallel and the perpendicular length scales was appropriately addressed, so that the FLR corrections have naturally emerged from our calculations. 

\subsection{Simplification of the system of equations}
Eq.\ref{normaliz_system_bis} provides a system of equations describing a plasma in the presence of an island whose width $w$ is larger than the ion Larmor radius $\rho_i$, so that the FLR corrections enter only to order $\rho^2$. In this approximation, we can reasonably assume $\alpha^2=O(1)$ \cite{fitzpatrick2009effect,fitzpatrick2012spontaneous}. If the ordering $\rho^2\ll 1$ holds, we can expand the fields in the following way:
\begin{equation}
\phi=\phi_0+\rho^2\phi_1+O(\rho^4)
\end{equation}
In the following calculations we are going to use the constant-$\psi$ approximation, which holds as long as $|\Delta'w|,\delta_e\ll1$, where $\Delta'$ is the tearing mode stability parameter and $\delta_e$ was defined previously. If this condition holds, the magnetic flux function takes the form:
\begin{equation}
\psi(x,y)=\frac{x^2}{2}+\cos y
\label{constant_psi}
\end{equation}
Eq.\ref{constant_psi} describes a magnetic island centered in $x=0$, with the O-point in $y=0$. The region inside the separatrix corresponds to $-1<\psi<1$ and the region outside the separatrix corresponds to $\psi\geq 1$.
From now on, the magnetic flux function $\psi$ will no longer be an unknown and, wherever possible, we will express the fields as functions of $\psi$ or its derivatives. For consistency with the results of Fitzpatrick \cite{fitzpatrick2009effect,fitzpatrick2012spontaneous}, we assume the zero-order fields to be flux functions, so that the first-order fields are going to be the lowest order FLR corrections. We define the following functions:
\begin{equation}
M\equiv\frac{d\phi_0}{d\psi},\hspace{10mm}L\equiv\frac{dn_0}{d\psi},\hspace{10mm}V_0'\equiv\frac{dV_0}{d\psi}
\end{equation}
The dissipative terms represented by the resistivity and the neoclassical viscosity are generally small, so that we can neglect them in first instance. We introduce the flux-surface average operation, which is defined as \cite{fitzpatrick2009effect}:
\begin{equation}
\left\langle f(\sigma,\psi,y)\right\rangle\equiv \left\{\begin{array}{ccc}
\frac{1}{2\pi}\oint dy\frac{f(\sigma,\psi,y)}{\sqrt{2(\psi-\cos y)}}\hspace{12mm}(\psi\geq 1)\\
\frac{1}{2\pi}\sum_{\sigma}\int_{-y_0}^{y_0}dy\frac{f(\sigma,\psi,y)}{\sqrt{2(\psi-\cos y)}}\hspace{1mm}(-1< \psi<1)
\end{array}\right.
\label{flux_surf_av}
\end{equation}
where $\sigma=\sign(x)$ and $y_0=\cos^{-1}\psi$. The flux surface average is the annihilator of the parallel gradient, so that every term in the form $[A,\psi]$ in the equations is deleted by this operator. By using the small-Larmor-radius expansion, we can find explicit expressions of the first-order fields in terms of the zero-order quantities $M$, $L$ and $V_0'$. To find the zero-order fields we need to introduce a second ordering which involves the transport coefficients. This new ordering assumes that the first order fields are as small as the transport coefficients, which are in turn much smaller than the FLR parameter $\rho^2$.\\
To recover the correct form of the equations \cite{fitzpatrick2009effect}, we introduce a phenomenological perpendicular viscosity $\mu$ and a diffusion coefficient $D$ (which is related to resistivity through the parallel compressibility \cite{fitzpatrick2005two}). The fields appearing in the final equations obey the following boundary conditions for $x\hspace{1mm}\rightarrow\hspace{1mm}\infty$ \cite{fitzpatrick2009effect}:
\begin{equation}\begin{split}
&n\hspace{2mm}\rightarrow\hspace{2mm}x\\&
\phi\hspace{2mm}\rightarrow\hspace{2mm}xV_p\\&
V\hspace{2mm}\rightarrow\hspace{2mm}V_{\infty}
\label{bound_cond}
\end{split}\end{equation}
The first condition means that the density gradient becomes constant far from the island. The gradient of the electrostatic potential tends to a constant value which is the electric field induced by the island rotation. The asymptotic velocity $V_{\infty}$ is determined by the neoclassical theory.

\section{Torque balance}
\label{sec4}
The linear stability index $\Delta'$ comes from the equilibrium current which causes the mode to be unstable, but every other contributions to the current affect the mode growth. It can be easily shown \cite{fitzpatrick2005effect} that, with the choice Eq.\ref{constant_psi} for the magnetic flux function, the contributions to the mode growth can be parametrized by this quantity :
\begin{equation}
J_c=4\int_{-1}^{+\infty}\left\langle J\cos y\right\rangle d\psi
\label{cos_contribution}
\end{equation}
where the angular brackets represent the flux-surface average operation. Eq.\ref{cos_contribution} means that the only currents that contribute to the mode growth are those which have the $\cos y$-symmetry. Analogously, there is a similar expression parametrizing the contributions to the torque which is exerted on the island by external currents:
\begin{equation}
J_s=4\int_{-1}^{+\infty}\left\langle J\sin y\right\rangle d\psi
\label{sin_contribution}
\end{equation}
Eq.\ref{sin_contribution} means that the only currents that contribute to the torque on the magnetic island are those which have the $\sin y$-symmetry. By solving the lowest-order vorticity equation:
\begin{equation}
[J_0^{(0)},\psi]-[\phi_0^{(0)},\partial^2_x(\phi_0^{(0)}+\tau/2n_0^{(0)})]-\tau/2[n_0^{(0)},\partial^2_x\phi_0^{(0)}]=0
\end{equation}
we find out that the solution for the current is:
\begin{equation}
J_0^{(0)}=\left(M'\left(M+\frac{\tau}{2}L\right)+\frac{\tau}{2}ML'\right)\widetilde{x^2}=\frac{1}{2}[M(M+\tau L)]'\widetilde{x^2}
\label{low_ord_curr}
\end{equation}
Since $\widetilde{x^2}=x^2-\left\langle x^2\right\rangle$, this term doesn't contribute to the torque. To find the lowest-order contribution to the torque, we have to consider the following first-order vorticity equation:
\begin{equation}\begin{split}
&[J^{(1)}_0,\psi]+[\phi^{(1)}_0,\partial^2_x(\phi^{(0)}_0+\tau/2n^{(0)}_0)]+\tau/2[n^{(1)}_0,\partial^2_x\phi^{(0)}_0]+[\phi^{(0)}_0,\partial^2_x(\phi^{(1)}_0+\tau/2n^{(1)}_0)]+\tau/2[n^{(0)}_0,\partial^2_x\phi^{(1)}_0]+\\&+\mu\partial^4_x(\phi^{(0)}_0+\tau/2 n^{(0)}_0)-\nu_{\perp}\partial^2_x[\phi^{(0)}_0+\tau n^{(0)}_0(1-c_{\perp})]+\nu_{\theta}\partial_x\{V^{(0)}_0-\partial_x[\phi^{(0)}_0+\tau n^{(0)}_0(1-c_{\theta})]\}=0
\label{first_ord_curr}
\end{split}\end{equation}
It follows by just performing the calculations and applying the boundary conditions, that the following identity holds:
\begin{equation}
\int_{-1}^{+\infty}\left\langle [J,\psi]x\right\rangle d\psi=-\int_{-1}^{+\infty}\left\langle J\sin y\right\rangle d\psi
\label{torque_identity_useful}
\end{equation}
Eq.\ref{torque_identity_useful} enables us to compute the lowest order contribution to the torque by just multiplying Eq.\ref{first_ord_curr} by $x$, solving it for $[J^{(1)}_0,\psi]x$ and operating on it with the flux-surface average and the $\psi$-integration. For an isolated island, which is not interacting with an external electromagnetic field, the total torque is zero. By doing this and applying again the boundary conditions, the torque-balance condition becomes:
\begin{equation}
\int_{-1}^{+\infty}d\psi \left\{\nu_{\theta}[(V_0-V_{\infty}+V_p+\tau(1-c_{\theta}))\left\langle 1\right\rangle+(M+\tau L(1-c_{\theta}))]+\nu_{\perp}[(V_p+\tau(1-c_{\perp}))\left\langle 1\right\rangle+(M+\tau L(1-c_{\perp}))]\right\}=0
\label{torque_balance}
\end{equation}
By using the system of equations for the fields we have deduced above, with their boundary conditions, together with the torque balance condition, we can find the phase velocity of the island in the following way: we first choose a value for the phase velocity $V_p$ and we solve the differential equations for the fields by the shooting method, we substitute these solutions in the torque balance condition and we find a new value for $V_p$, we use this new value in the equations again and we iterate until convergence is reached.\\
$V_{\infty}$ represents the velocity of the plasma far from the island, which depends on the damping effects. We will see in the following sections that a solution can be found in two different collisionality regime, namely the weak damping and the intermediate damping regimes.
\section{Weak damping regime}
\label{sec5}
In the weak-damping regime, the following ordering holds:
\begin{equation}
1\gg D,\mu,\eta,\nu_{\theta}\gg \nu_{\perp}
\end{equation}
In this case, we can disregard the terms where the products between the FLR parameter $\rho^2$ and the transport coefficients appear, as well as the perpendicular damping coefficient $\nu_{\perp}$. Also the product $(\epsilon/q)^2\nu_{\theta}$ is small and can be neglected. In the weak-damping regime, the equations become:
\begin{equation}\begin{split}
&D\left\langle \partial^2_x n^{(0)}_0\right\rangle+\rho^2/2\left\langle[x^2M',n^{(1)}_0]\right\rangle=0\\&
\mu\left\langle \partial^2_x V^{(0)}_0\right\rangle+\rho^2\left\{\left\langle [\phi^{(0)}_1,V^{(1)}_0]\right\rangle+\left\langle [\phi^{(1)}_0,V^{(0)}_1]\right\rangle+1/2\left\langle [x^2M',V^{(1)}_0]\right\rangle\right\}=0\\&
\mu\left\langle \partial^4_x (\phi^{(0)}_0+\tau n^{(0)}_0)\right\rangle+\left\langle [\phi^{(1)}_0,x^2(M'+\tau/2L')]\right\rangle-\nu_{\theta}\left\langle \partial_x\{V^{(0)}_0-\partial_x[\phi^{(0)}_0+\tau n^{(0)}_0(1-c_{\theta})]\}\right\rangle+\\&+\tau/2\left\langle [n^{(1)}_0,x^2M']\right\rangle+\rho^2\left\{\left\langle [\phi^{(1)}_0,\partial^2_x(\phi^{(0)}_1+\tau/2n^{(0)}_1)]\right\rangle+\left\langle [\phi^{(1)}_0,x^4(M'''+\tau/2L''')+6x^2(M''+\tau/2L'')]\right\rangle+\right.\\&\left.+\tau/2\left\langle [n^{(1)}_0,\partial^2_x\phi^{(0)}_1]\right\rangle+\tau/4\left\langle [n^{(1)}_0,x^4L'''+6x^2L'']\right\rangle\right\}=0
\label{transport_ord_weak}
\end{split}\end{equation}
Eqs.\ref{transport_ord_weak} neglect the FLR corrections to the neoclassical flow damping. After many mathematical steps, the system of equations reduces to  
\begin{equation}\begin{split}
&\frac{d}{d\psi}\left( L\left\langle x^2\right\rangle\right)=0,\hspace{10mm}\frac{d}{d\psi}\left(\left\langle x^2\right\rangle\frac{dV_0}{d\psi}\right)=0\\&
\frac{d}{d\psi}\left[\left\langle x^4\right\rangle\frac{d}{d\psi}\left(M+\tau L\right)\right]-\left[\left(1+\frac{\tau}{2}\right)\frac{M'}{H}+\frac{\tau}{2}\frac{L'}{H}\right]\left(\frac{D}{\mu}ML'+V_0''\right)\left\langle \widetilde{x^2}\widetilde{x^2}\right\rangle-\\&-\frac{\nu_{\theta}}{\mu}\left\{(V_0-V_{\infty}+V_p+\tau(1-c_{\theta}))+[M+\tau L(1-c_{\theta})]\left\langle x^2\right\rangle\right\}=0
\label{weak_damp_syst}
\end{split}\end{equation}
where we used $V_0=V^{(0)}_0$. Note that the solution of the first equation, compatible with the boundary condition $V\hspace{1mm}\rightarrow\hspace{1mm}V_{\infty}$, is \cite{fitzpatrick2009effect}:
\begin{equation}
L=-1/\left\langle x^2\right\rangle
\label{sol_for_L}
\end{equation}
The solution for $V_0$, compatibly with the boundary condition $V_0\hspace{1mm}\rightarrow\hspace{1mm}V_{\infty}$, is $V_0=V_{\infty}$.\\
We can neglect the island-induced flow damping in the torque-balance condition Eq.\ref{torque_balance}, which becomes:
\begin{equation}
\int_{-1}^{+\infty}d\psi\left\{(V_0-V_{\infty}+V_p+\tau(1-c_{\theta}))\left\langle 1\right\rangle+(M+\tau L(1-c_{\theta}))\right\}=0
\label{torque_balance_weak1}
\end{equation}
Now we consider the fact that $\phi_0$ and $n_0$ are flux functions, so that they must have the same $x$-symmetry of the flux function $\psi$, which is an even function. However, for the tearing symmetry, both $\phi_0$ and $n_0$ are even in respect to $x$. The only way to solve this contradiction is by imposing that they must be zero inside the separatrix \cite{fitzpatrick2009effect}, that is for $-1<\psi<1$. By using this result in the parallel momentum equation, we find out that $V_0=-V_p$ inside the separatrix. Furthermore, the quantity $c_{\theta}$ which drives the intrinsic poloidal rotation, depends on the temperature gradient and it is thus zero inside the separatrix. Outside the separatrix, instead, the system of equations Eq.\ref{weak_damp_syst} hold. The first two equations bring to the solutions we have already seen $L=-1/\left\langle x^2\right\rangle$ and $V_0=V_{\infty}$. Since $V_0$ is a constant, the relation $V_0=V_{\infty}$ holds in all the region $1<\psi<+\infty$. By imposing that $V_0$ is continuous across the separatrix, we also find that $V_p=-V_{\infty}$. By putting these results in the torque balance condition Eq.\ref{torque_balance_weak1}, we find an equation for $V_p$, whose solution is
\begin{equation}
V_p=-\tau\left[1-\frac{c_{\theta}}{I_1}(I_2-I_3)\right]-\frac{1}{I_1}\int_1^{+\infty}d\psi(M+\tau L)
\label{phase_vel_weak}
\end{equation}
where we introduced the quantities $I_1=\int_{-1}^{+\infty}d\psi \left\langle 1\right\rangle$, $I_2=\int_1^{+\infty}d\psi \left\langle 1\right\rangle$ and $I_3=-\int_1^{+\infty}d\psi L$. The remaining unknown function $M$ must be determined by solving the following equation:
\begin{equation}
\frac{d}{d\psi}\left[\left\langle x^4\right\rangle\frac{d}{d\psi}\left(M+\tau L\right)\right]-\frac{D}{\mu}\left[\left(1+\frac{\tau}{2}\right)\frac{M'}{H}+\frac{\tau}{2}\frac{L'}{H}\right] ML'\left\langle \widetilde{x^2}\widetilde{x^2}\right\rangle-\frac{\nu_{\theta}}{\mu}(V_p+M\left\langle x^2\right\rangle)=0
\label{equ_M_weak}
\end{equation}
where $H=M(L-M)+\alpha^2(1+\tau)$. By solving Eq.\ref{equ_M_weak} with the boundary condition $M\hspace{1mm}\rightarrow\hspace{1mm}V_p/\sqrt{2\psi}$ and computing $V_p$ by using Eq.\ref{phase_vel_weak} iteratively, we can obtain the radial profile of $M$ and the phase velocity $V_p$.

\section{Intermediate damping regime}
\label{sec6}
In the intermediate-damping regime, the following ordering holds:
\begin{equation}
1\gg \nu_{\theta}\gg D,\mu,\eta,\nu_{\perp}
\label{interm_damp}
\end{equation}
In this case, we have to keep the terms where the product between $\rho^2$ and the poloidal damping coefficient $\nu_{\theta}$ appear, together with the non-axisymmetric contributions $\nu_{\perp}$. By neglecting again the products $\rho^2 D$, $\rho^2 \mu$ and $\rho^2\nu_{\theta}(\epsilon/q)^2$, the set of surface-averaged equations becomes:
\begin{equation}\begin{split}
&\frac{d}{d\psi}\left(\left\langle x^2\right\rangle\frac{dV_0}{d\psi}\right)-\frac{\nu_{\theta}}{\mu}\left(\frac{\epsilon}{q}\right)^2\left\{(V_0+V_p)\left\langle 1\right\rangle+[M+\tau L(1-c_{\theta})]\right\}=0\\&
\frac{d}{d\psi}\left[\left\langle x^4\right\rangle\frac{d}{d\psi}\left(M+\tau L\right)\right]-\frac{\nu_{\theta}}{\mu}(V_0-V_{\infty}+V_p+M\left\langle x^2\right\rangle)-\frac{\nu_{\perp}}{\mu}(V_p+M\left\langle x^2\right\rangle)-\\&-\left[\left(1+\frac{\tau}{2}\right)\frac{M'}{H}+\frac{\tau}{2}\frac{L'}{H}\right]\left[\left(\frac{D}{\mu}ML'+V_0''\right)\left\langle \widetilde{x^2}\widetilde{x^2}\right\rangle+\frac{\nu_{\theta}}{\mu}\left(\frac{\epsilon}{q}\right)^2[M+\tau L(1-c_{\theta})]\left\langle \widetilde{x^2}\tilde{x}\right\rangle\right]+\\&+\rho^2\frac{\nu_{\theta}}{\mu}\left\langle x\frac{d}{d\psi}\left[\frac{A}{\tau H}\widetilde{x^2}-\left(\frac{1}{2}-\tau c_2\right)M'x^2-\frac{B}{H}\widetilde{x^2}(1-c_{\theta})+\tau c_1L'x^2\right]\right\rangle=0
\label{interm_damp_system}
\end{split}\end{equation}
where $H\equiv V_0'+M(L-M)+\alpha^2(1+\tau)$, $A\equiv M^2\left(M'+\tau/2L'\right)+\tau/2MLM'-\alpha^2(1+\tau)\left(M'+\tau/2L'\right)+\tau/2(V_0'M'+\alpha^2L')-M'$ and $B\equiv A+H\left(M'+\tau/2L'\right)$. In the torque balance, we have to include the island-induced damping terms proportional to $\nu_{\perp}$. In the internal region, the same considerations hold as before. In the external region, however, we can impose the further condition that the plasma velocity tends to the intrinsic poloidal velocity far from the island. This condition is equivalent to imposing that the poloidal flow damping tends to zero far from the island, that is:
\begin{equation}
\nu_{\theta}\left\{V+V_p-\partial_x[\phi+\tau n(1-c_{\theta})]\right\}\hspace{2mm}\rightarrow\hspace{2mm}0
\label{pol_damp_zero}
\end{equation}
By imposing the boundary conditions, Eq.\ref{pol_damp_zero} becomes $V_{\infty}=\tau(1-c_{\theta})$. By using this result, the torque-balance condition Eq.\ref{torque_balance} becomes an equation for $V_p$, whose solution is
\begin{equation}
V_p=-\tau\left[1-\frac{\nu_{\theta}c_{\theta}+\nu_{\perp}c_{\perp}}{(\nu_{\theta}+\nu_{\perp})I_1}\left(I_2-I_3\right)\right]-\frac{\nu_{\theta}}{(\nu_{\theta}+\nu_{\perp})I_1}\int_1^{+\infty}d\psi\left\langle 1\right\rangle\left(\tau(1-c_{\theta})-V_0\right)-\frac{1}{I_1}\int_1^{+\infty}d\psi(M+\tau L)
\label{phase_vel_interm}
\end{equation}
In the limit $\nu_{\perp}\rightarrow 0$ and $V_0=\tau(1-c_{\theta})$, which is the case of the weak-damping regime, Eq.\ref{phase_vel_interm} reduces to Eq.\ref{phase_vel_weak}. The remaining unknown functions $V_0$ and $M$ must be determined by solving the system Eqs.\ref{interm_damp_system}. By taking a few more steps, the term containing the FLR corrections to the poloidal flow damping can be written more explicitly, so that Eqs.\ref{interm_damp_system} become:
\begin{equation}\begin{split}
&\frac{d}{d\psi}\left(\left\langle x^2\right\rangle\frac{dV_0}{d\psi}\right)-\frac{\nu_{\theta}}{\mu}\left(\frac{\epsilon}{q}\right)^2\left\{(V_0+V_p)\left\langle 1\right\rangle+[M+\tau L(1-c_{\theta})]\right\}=0\\&
\frac{d}{d\psi}\left[\left\langle x^4\right\rangle\frac{d}{d\psi}\left(M+\tau L\right)\right]-\frac{\nu_{\theta}}{\mu}(V_0+V_p-\tau(1-c_{\theta})+M\left\langle x^2\right\rangle)-\frac{\nu_{\perp}}{\mu}(V_p+M\left\langle x^2\right\rangle)-\\&-\left[\left(1+\frac{\tau}{2}\right)\frac{M'}{H}+\frac{\tau}{2}\frac{L'}{H}\right]\left[\left(\frac{D}{\mu}ML'+V_0''\right)\left\langle \widetilde{x^2}\widetilde{x^2}\right\rangle+\frac{\nu_{\theta}}{\mu}\left(\frac{\epsilon}{q}\right)^2[M+\tau L(1-c_{\theta})]\left\langle \widetilde{x^2}\tilde{x}\right\rangle\right]+\\&+\rho^2\frac{\nu_{\theta}}{\mu}\frac{d}{d\psi}\left[\left(\frac{1}{\tau H}[A+\tau B(1-c_{\theta})]\left(1-\left\langle 1\right\rangle\right)-\left(\frac{1}{2}-\tau c_2\right)M'+\tau c_1 L'\right)\left\langle x^2\right\rangle\right]=0
\label{equ_V0-M_interm}
\end{split}\end{equation}
where $H=V_0'+M(L-M)+\alpha^2(1+\tau)$, $A$ and $B$ have been defined above. Just as in the weak-damping case, by solving Eq.\ref{equ_V0-M_interm} with the boundary conditions $M\hspace{1mm}\rightarrow\hspace{1mm}V_p/\sqrt{2\psi}$, $V_0\hspace{1mm}\rightarrow\hspace{1mm}\tau(1-c_{\theta})$ and computing $V_p$ by using Eq.\ref{phase_vel_interm} iteratively, we can obtain the radial profiles of $M$ and $V_0$ and the phase velocity $V_p$.
\subsection{Further simplification}
Eqs.\ref{equ_V0-M_interm} can be further simplified by considering the limit of small-Larmor-radius, $\rho^2\rightarrow 0$, and the ordering Eq.\ref{interm_damp}. By using these simplifications, $V_0$ outside the separatrix can be deduced by solving the second equation of Eqs.\ref{equ_V0-M_interm}:
\begin{equation}
V_0=\tau(1-c_{\theta})-(V_p+M\left\langle x^2\right\rangle)\frac{\nu_{\theta}+\nu_{\perp}}{\nu_{\theta}}
\label{V_0}
\end{equation}
For the reasons explained above, $V_0=-V_p$ inside the separatrix. Substituing Eq.\ref{V_0} in Eq.\ref{torque_balance}
\begin{equation}
V_p=-\tau\left[1+\frac{(\nu_{\theta}+\nu_{\perp})I_2}{\nu_{\perp}(I_1-I_2)}-\frac{(\nu_{\theta}c_{\theta}+\nu_{\perp}c_{\perp})}{\nu_{\perp}(I_1-I_2)}(I_2-I_3)\right]-\frac{(\nu_{\theta}+\nu_{\perp})}{\nu_{\perp}(I_1-I_2)}\int_1^{+\infty}d\psi[M(1-\left\langle x^2\right\rangle\left\langle 1\right\rangle)+\tau L]
\label{vel_interm_bis}
\end{equation}
By substituing Eq.\ref{V_0} in Eqs.\ref{equ_V0-M_interm}, we get:
\begin{equation}\begin{split}
&\frac{\nu_{\theta}+\nu_{\perp}}{\nu_{\theta}}\frac{d}{d\psi}\left[\left\langle x^2\right\rangle\frac{d}{d\psi}\left(M\left\langle x^2\right\rangle\right)\right]-\frac{\nu_{\perp}}{\mu}\left(\frac{\epsilon}{q}\right)^2[V_p+M\left\langle x^2\right\rangle]\left\langle 1\right\rangle+\\&+\frac{\nu_{\theta}}{\mu}\left(\frac{\epsilon}{q}\right)^2\left\{M(1-\left\langle x^2\right\rangle\left\langle 1\right\rangle)+\tau(1-c_{\theta})(\left\langle 1\right\rangle+L)\right\}=0
\label{Eq_per_M}
\end{split}\end{equation}
By solving simultaneously Eq.\ref{vel_interm_bis} and Eq.\ref{Eq_per_M}, with the appropriate boundary conditions, we can obtain the radial profile of $M$ and the phase velocity $V_p$. The systems of equations Eq.\ref{phase_vel_weak}+Eq.\ref{equ_M_weak} and Eq.\ref{vel_interm_bis}+Eq.\ref{Eq_per_M} represent limit cases which can be easily solved numerically, and they can be both deduced from Eq.\ref{phase_vel_interm}+Eq.\ref{equ_V0-M_interm} under appropriate limits. To further simplify the calculations, we introduce the variable $k=\sqrt{(1-\psi)/2}$. Then we define the following quantities, $E=\left\langle x^2\right\rangle/(2k)$, $F=2k\left\langle 1\right\rangle$ and we define the variable $Q$ so as to absorb the factor $V_p+\tau(1-c_{\theta})$:
\begin{equation}
Q=\frac{\tau(1-c_{\theta})-2kEM}{\tau(1-c_{\theta})+V_p}
\end{equation}
By using these quantities, Eq.\ref{vel_interm_bis} and Eq.\ref{Eq_per_M} become:
\begin{equation}
V_p=-\tau\left[1+\frac{I_6}{I_4+I_5}(c_{\theta}-c_{\perp})-c_{\theta}\frac{I_5}{I_4+I_5}\right]
\label{vel_interm_biss}
\end{equation}
\begin{equation}\begin{split}
&\frac{\nu_{\theta}+\nu_{\perp}}{4\nu_{\theta}}\frac{d}{dk}\left[E\frac{dQ}{dk}\right]-\frac{\nu_{\perp}}{\mu}\left(\frac{\epsilon}{q}\right)^2F(Q-1)-\frac{\nu_{\theta}}{\mu}\left(\frac{\epsilon}{q}\right)^2Q\left(F-\frac{1}{E}\right)=0
\label{Eq_per_M_biss}
\end{split}\end{equation}
where $I_4=\int_0^1 2k\left\langle 1\right\rangle dk$, $I_5=(\nu_{\theta}+\nu_{\perp})/\nu_{\perp}\int_1^{+\infty}Q(F-1/E)dk$ and $I_6=\int_1^{+\infty}(F-1/E)dk$.
\section{Numerical results}
\label{sec7}
\begin{figure}[httb!]
  \centering
  \includegraphics[width=0.5\linewidth]{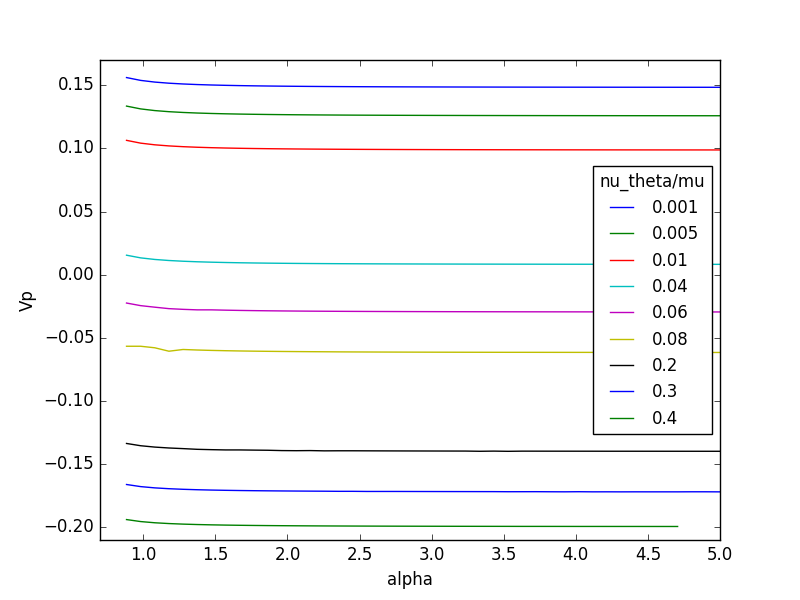}
  \caption{Island phase velocity $V_p$ versus the ion-acoustic parameter $\alpha$ for different choices of the poloidal flow damping parameter $\nu_{\theta}/\mu$}
  \label{fig:weak_damping}
\end{figure}

\begin{figure}[httb!]
  \centering
  \includegraphics[width=0.5\linewidth]{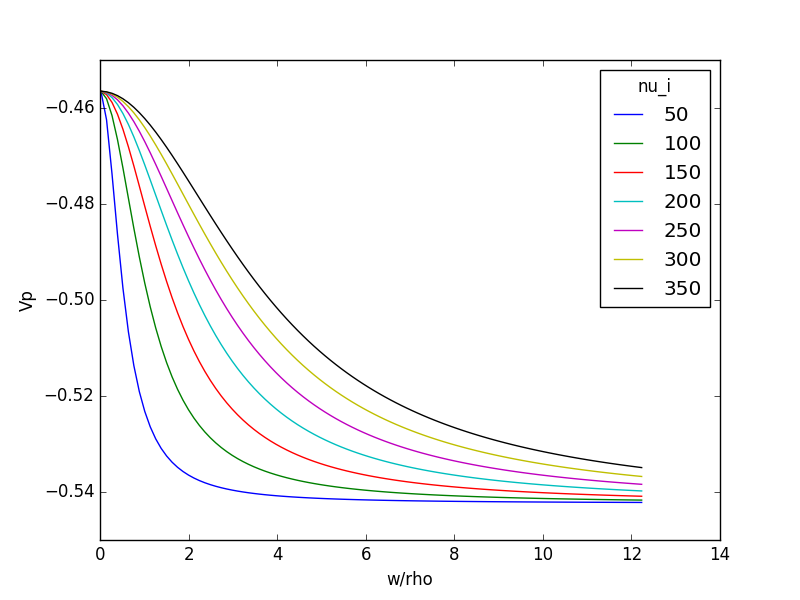}
  \caption{Island phase velocity $V_p$ versus the normalized island width $w/\rho_s$ for different choices of the collision frequency $\nu_i$ with $\eta_i=0.5$}
  \label{fig:interm_damping1}
\end{figure}

\begin{figure}[httb!]
  \centering
  \includegraphics[width=0.5\linewidth]{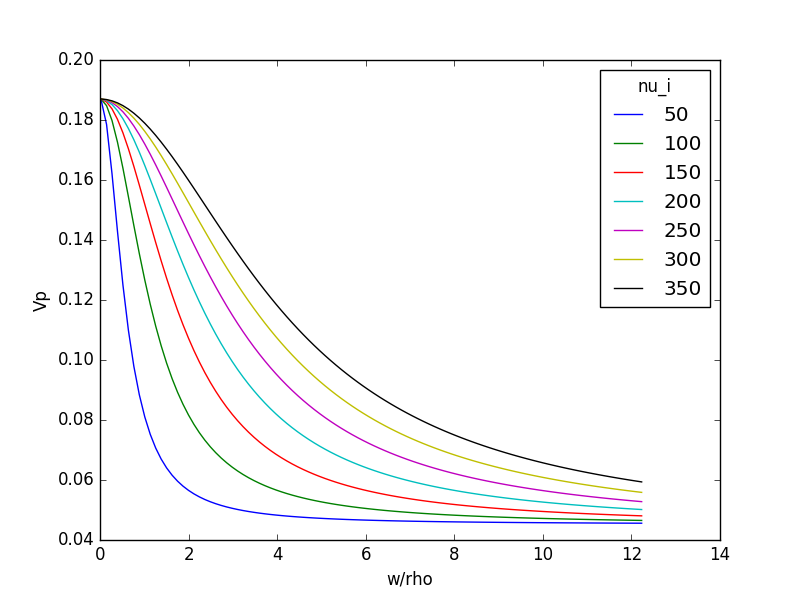}
  \caption{Island phase velocity $V_p$ versus the normalized island width $w/\rho_s$ for different choices of the collision frequency $\nu_i$ with $\eta_i=1$}
  \label{fig:interm_damping2}
\end{figure}

\begin{figure}[httb!]
  \centering
  \includegraphics[width=0.5\linewidth]{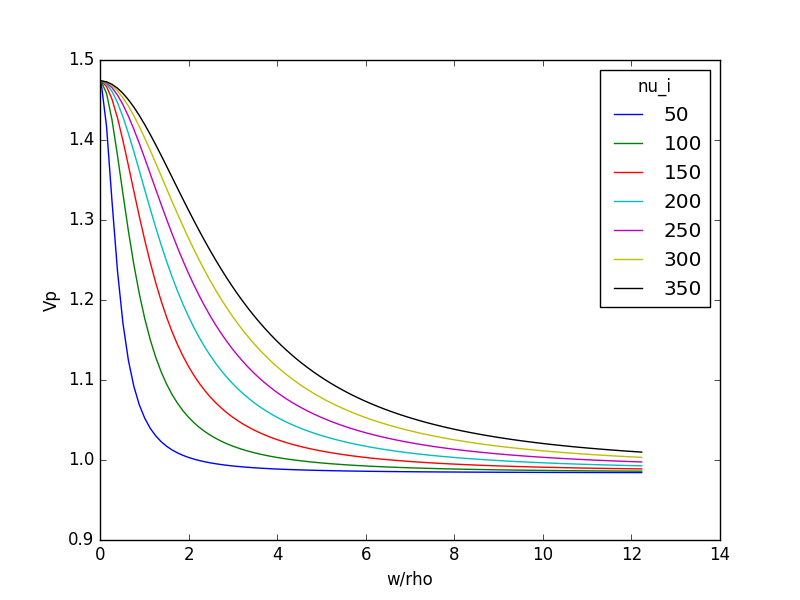}
  \caption{Island phase velocity $V_p$ versus the normalized island width $w/\rho_s$ for different choices of the collision frequency $\nu_i$ with $\eta_i=2$}
  \label{fig:interm_damping3}
\end{figure}

It is possible to determine the magnetic islands rotation velocity by numerically integrating the system of equations consisting in Eqs.\ref{phase_vel_weak},\ref{equ_M_weak} for the weak damping regime and Eqs.\ref{vel_interm_biss},\ref{Eq_per_M_biss} for the intermediate damping regime, in the limit of small Larmor radius. Unfortunately, the numerical integration of Eqs.\ref{phase_vel_interm},\ref{equ_V0-M_interm} still represents a challenge too difficult to solve. We leave the analysis of the more general case to a future work.\\
The weak damping regime is characterized by the parameters $\alpha$, which enters the denominator $H$, and $\nu_{\theta}/\mu$. $\alpha$ is proportional to the ratio between the island width $w$ and the ion-acoustic radius $\rho_s$, and it is a measure of the importance of the ion-acoustic waves on the flattening of the density profile inside the separatrix. $\nu_{\theta}/\mu$ depends on the plasma collisionality and measures the importance of the poloidal flow damping. The result of the numerical integration of Eqs.\ref{phase_vel_weak},\ref{equ_M_weak} for different values of $\alpha$ and $\nu_{\theta}/\mu$ is displayed in Fig.\ref{fig:weak_damping}.
The presence of the resonant denominator $H$ in Eq.\ref{equ_M_weak} prevents the solution from converging for the smaller values of $\alpha$, which correspond to the small island width limit. The interesting feature of Fig.\ref{fig:weak_damping} is the transition of the value of $V_p$ from positive values to negative values as the parameter $\nu_{\theta}/\mu$ is increased from values much smaller than one to values close to one. A positive phase velocity corresponds to an island rotating in the direction of the electron fluid, while a negative value corresponds to the direction of the ion fluid. The hypotheses of zero equilibrium electric field means that the $\boldsymbol E \wedge \boldsymbol B$ drift has been subtracted from the plasma velocity.\\
The intermediate damping regime, in the limit of small Larmor radius, is characterized by the parameters $w/\rho_s$, which enters the perpendicular damping coefficient, and $\nu_i$, which enters both the poloidal and the perpendicular damping coefficients. From Eq.\ref{vel_interm_biss}, it is evident that the island phase velocity $V_p$ is determined by the neoclassical velocities $c_{\theta}$ and $c_{\perp}$, which are proportional to the radial temperature gradient through the parameter $\eta_i=L_n/L_T$. The result of the numerical integration of Eqs.\ref{phase_vel_weak},\ref{equ_M_weak} for different values of $w/\rho_s$ and $\nu_i$ is displayed in Figs.\ref{fig:interm_damping1},\ref{fig:interm_damping2},\ref{fig:interm_damping3} for the choices of $\eta_i=0.5$, $\eta_i=1$ and $\eta_i=2$.

Because of the absence of a resonant denominator, the solution converges even in the limit of small island width. However, the equations we integrate are valid only in the limit of small Larmor radius, so that the results lose validity for $w/\rho_s<1$. The interesting feature of these pictures is the transition of the value of $V_p$ from negative values to positive values as the parameter $\eta_i$ is increased from values less than one to values larger than one. Note that $L_n$ and $L_T$ are of the same order in realistic tokamak plasmas, so that very large values and very small values of $\eta$ are unrealistic. The different slopes of the curves corresponding to the different values of $\nu_i$ show that, the smaller the collisionality, the more effective the neoclassical flow damping is in relaxing the island velocity towards the neoclassical value, which is determined by the parameters $c_{\theta}$ and $c_{\perp}$. Note that the numerical integration of Eqs.\ref{phase_vel_weak},\ref{equ_M_weak} for the weak damping regime was performed with the choice $\eta_i=1$, but the results are not significantly affected by the choice of $\eta_i$. In all our integrations we chose $\tau=1$.
\section{Conclusions}
\label{sec8}
In this paper we addressed the issue of determining the phase velocity of a chain of freely rotating magnetic islands by using a four field gyrofluid system of equations which includes the neoclassical flow damping effects and the lowest order FLR corrections. To do that, we first solved the gyrokinetic equation under some simplifying hypotheses and we computed the FLR corrections to the poloidal flow damping. Then we deduced a four field gyrofluid model by starting from a set of gyrofluid equations and we closed it by using a simplified form for the divergence of the stress tensor, which provides the neoclassical flow damping effects. By following the method described by Fitzpatrick \& Waelbroek \cite{fitzpatrick2005two,fitzpatrick2006influence,fitzpatrick2008drift,fitzpatrick2009effect}, we managed to obtain a system of equations whose solution provides the islands rotation velocity consistently with the fields radial profiles close to the resonant surface. We applied this system of equations to the investigation of two collisionality regimes, namely the weak damping regime and the intermediate damping regime. In the second case, which corresponds to the low collisionality regime, an additional term, containing the lowest order FLR corrections to the poloidal flow damping, appeared in the equations. The numerical integration of Eqs.\ref{phase_vel_weak},\ref{equ_M_weak} in the weak damping regime shows that the island phase velocity moves from positive values to negative values as the poloidal damping parameter $\nu_{\theta}/\mu$ is increased from values much smaller than one to values close to one. A positive phase velocity is associated with a magnetic island rotating in the direction of the electron fluid, while negative values means that the island rotates in the direction of the ions. The numerical integration of Eqs.\ref{vel_interm_biss},\ref{Eq_per_M_biss} in the intermediate damping regime shows that the phase velocity $V_p$ moves from negative values to positive values as the parameter $\eta_i$ is increased from values less than one to values larger than one. These results are in agreement with what was already known about the subject, but they are valid only within the limitations of their hypotheses. Unfortunately, the numerical integration of Eqs.\ref{phase_vel_interm},\ref{equ_V0-M_interm} still represents a challenge too difficult to solve. We leave the analysis of the more general case to a future work. In the case of large, saturated islands, we expect $\rho^2$ to be small, so that the FLR corrections we found should be small as well. However, if the islands are not much smaller than the ion acoustic radius, or if we are in the case of a high temperature plasma, the product $\rho^2\nu_{\theta}/\mu$ multiplying the additional term in Eq.\ref{equ_V0-M_interm} might be comparable with the other terms entering the equations. The procedure we used to deduce the final equations and to numerically solve them is thorughly described in \cite{fitzpatrick2008drift,fitzpatrick2009effect}. Although this procedure is based on the assumption that the island width is much larger than the ion-acoustic radius, the so called sonic regime, the extension of this approach to the hypersonic regime would require a few changes in the initial hypotheses. In particular, when we deal with hypersonic islands, the hypothesis that the lowest order fields are flux functions is no longer valid, and the effect of the drift-acoustic waves must be kept into account to determine the radial profiles of density and electrostatic potential inside the separatrix.

\bibliographystyle{ieeetr}

\bibliography{biblio_compless}

\end{document}